\documentclass[
  aps,
  english,
  reprint,
  floatfix,
  superscriptaddress
]{revtex4-1}

\usepackage[T1]{fontenc}
\usepackage[utf8]{inputenc}
\usepackage{amsmath}
\usepackage{amssymb}
\usepackage{graphicx}
\usepackage[x11names]{xcolor}
\usepackage[
  allcolors=Blue3,
  colorlinks,
  pdftex,
  pdfauthor={Manu Mannattil, Ambrish Pandey, Mahendra K. Verma, and Sagar Chakraborty},
  pdftitle={On the Applicability of Low-Dimensional Models for Convective Flow Reversals at Extreme Prandtl Numbers},
  pdfsubject={fluid dynamics, chaos},
  pdfkeywords={turbulence, convection, nonlinear time series analysis}
]{hyperref}
\usepackage[helvratio=0.92]{newtxtext}
\usepackage[amssymbols,bigdelims,cmintegrals]{newtxmath}

\DeclareSymbolFont{largesymbolsCM}{OMX}{cmex}{m}{n}

\let\sum\relax
\DeclareMathSymbol{\sum}{\mathop}{largesymbolsCM}{"50}

\usepackage{bm}
\renewcommand{\mathbf}{\bm}

\allowdisplaybreaks
\begin{document}

\title{On the Applicability of Low-Dimensional Models for Convective Flow Reversals\\at Extreme Prandtl Numbers}

\author{Manu Mannattil}
\thanks{Present address: Department of Physics, Syracuse University, Syracuse, New York 13244, USA; \href{mailto:mmannatt@syr.edu}{mmannatt@syr.edu}}
\affiliation{
  Department of Physics,
  Indian Institute of Technology Kanpur,
  Uttar Pradesh 208016, India
}
\author{Ambrish Pandey}
\email{pambrish@iitk.ac.in}
\affiliation{
  Department of Physics,
  Indian Institute of Technology Kanpur,
  Uttar Pradesh 208016, India
}
\author{Mahendra K. Verma}
\email{mkv@iitk.ac.in}
\affiliation{
  Department of Physics,
  Indian Institute of Technology Kanpur,
  Uttar Pradesh 208016, India
}
\affiliation{
  Mechanics and Applied Mathematics Group,
  Indian Institute of Technology Kanpur,
  Uttar Pradesh 208016, India
}
\author{Sagar Chakraborty}
\email{sagarc@iitk.ac.in}
\affiliation{
  Department of Physics,
  Indian Institute of Technology Kanpur,
  Uttar Pradesh 208016, India
}
\affiliation{
  Mechanics and Applied Mathematics Group,
  Indian Institute of Technology Kanpur,
  Uttar Pradesh 208016, India
}


\begin{abstract}
  Constructing simpler models, either stochastic or deterministic, for exploring the phenomenon of flow reversals in fluid systems is in vogue across disciplines.
  Using direct numerical simulations and nonlinear time series analysis, we illustrate that the basic nature of flow reversals in convecting fluids can depend on the dimensionless parameters describing the system.
  Specifically, we find evidence of low-dimensional determinism in flow reversals occurring at zero Prandtl number, whereas we fail to find such signatures for reversals at infinite Prandtl number.
  Thus, even in a single system, as one varies the system parameters, one can encounter reversals that are fundamentally different in nature.
  Consequently, we conclude that a single general low-dimensional deterministic model cannot faithfully characterize flow reversals for every set of parameter values.
\end{abstract}

\maketitle


\section{Introduction}

Complete scientific understanding of any real-life phenomenon in its entirety and in its minutest possible mathematical detail is almost always impossible.
The standard approach towards solving many problems in physics, therefore, is to understand them through tractable models that are conceived after simplifying the physics to its essential minimum, so that the dominant reason behind the phenomena are revealed within experimental error.
It is for this reason that simpler, low-dimensional models (LDMs) are of immense practical importance to physicists and to researchers in other scientific disciplines.

Consider, for example, the reversal dynamics of thermal convection.
Experiments~\cite{niemela01, sreeni02, brown05,xi06, sugiyama10} and numerical simulations~\cite{sugiyama10,chandra11,chandra13,ni15} reveal seemingly random reversals of the vertical velocity field at points near the lateral walls of a container with thermally-driven fluid.
By reversal, one means that the velocity field at a point in the fluid flips its direction to become antiparallel to its earlier direction.
Such reversals are broadly called \emph{field reversals}, or in the case of fluids, \emph{flow reversals}.
A related phenomenon is geologists' discovery that the Earth's magnetic field has reversed several times in the past, with the interval between two consecutive field reversals being randomly distributed~\cite{glatz95,dormy07}.
Similarly, observations of the Sun show that its magnetic field reverses its direction every 11 years, with the transitions often affecting space weather~\cite{petrovay10}.
Such reversals can often show complex behavior, such as the appearance of multipolar magnetic field structures.
Reversals thought to be closely related to flow reversals and magnetic field reversals have also been reported in other natural systems, e.g., in oceanic thermohaline circulation-affected paleoclimate~\cite{cronin99}.

The phenomenon of flow reversals is complex, and thus, as motivated earlier, to gain a simplified understanding of their dynamics, researchers have constructed LDMs~\cite{araujo05,gissinger10,podvin15}.
\citet{araujo05} derived three nonlinearly-coupled ordinary differential equations representing large-scale temperature Fourier modes.
Numerical simulations of these equations exhibit random reversals in the temperature field.
\citet{gissinger10} constructed a model containing three ordinary differential equations that represent the dipolar and quadrupolar magnetic field, and the large-scale velocity field.
This model exhibits reversals similar to that observed in the dynamo field generated during the turbulent flow of liquid sodium.
\citet{podvin15} used proper orthogonal decomposition to construct an LDM that captures the reversals in two-dimensional thermal convection.
\begin{figure*}
  \centering
  \includegraphics[]{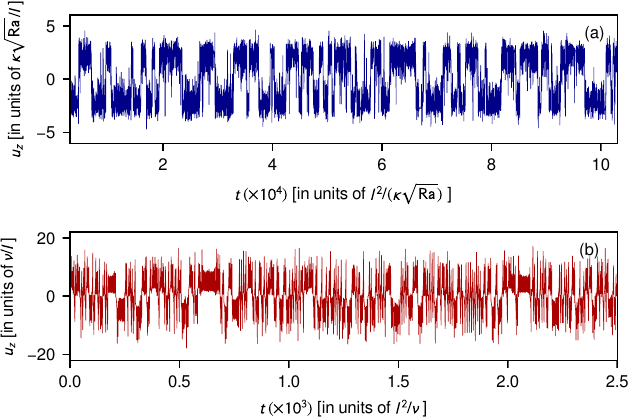}
  \caption{Irregular time series of the vertical real space velocities $u_z(t)$ for reversals at (a)~infinite Pr and (b)~zero Pr.  Further details about the time and velocity units as well as the Rayleigh--B\'enard convection generating these reversals have been given in the main text.\label{fig:series}}
\end{figure*}

Researchers have also highlighted the similarities between the two quasi-steady (positive and negative) states of large-scale circulation with bistability.
Various proposed models employ noise to switch the system between the two stable states that leads to randomization of the flow reversals.
One such model is by \citet{sreeni02} who showed similarities between the statistics of reversal times and a noisy overdamped bistable oscillator.
\citet{brown08} modeled flow reversals in a cylindrical geometry using stochastic differential equations that represent the evolution of the azimuthal orientation and azimuthal temperature amplitudes.
\citet{petrelis09} modeled the reversals in a liquid-sodium dynamo as a saddle-node bifurcation in the amplitude equation; they add noise in the system for bringing in stochasticity in the dynamo reversals.
In several models for field reversals, higher Fourier modes representing small-scale structures provide stochasticity to the large-scale flow.
One such model is by Benzi~\cite{benzi05} who explained random reversals using a shell model for hydrodynamic turbulence.
Here, the first mode of the model is compared with the large-scale circulation of convection, whereas the higher modes provide randomness.
\citet{benzi08} employed the Ginzburg--Landau equation to study reversals; they relate the large-scale flow to the mean value of the field of the Ginzburg--Landau equation.

It should be emphasized that none of the aforementioned models, low-dimensional or stochastic, capture all the features of the systems, e.g., neither the LDM provided by~\citet{araujo05} nor the stochastic model of~\citet{benzi05} explain all the features of flow reversals seen in convection experiments~\cite{brown05,xi06}.
Beyond all this, whether one should always prefer low-dimensional deterministic models over stochastic ones is not clear.
Our investigation provides an insight into this aspect of the problem.

In Fig.~\ref{fig:series} we illustrate the phenomenon of flow reversals by plotting numerically generated time series of the vertical velocity component at a point in the bulk of convecting fluid characterized by, among other parameters, the Prandtl number (Pr)---the ratio of kinematic viscosity to thermal diffusivity of the fluid.
As described in more details in Section~\ref{sec:numerics}, these time series have been generated using very high-dimensional direct numerical simulations of the Navier--Stokes equations adapted for the convection problem.
Originating from first principles, the Navier--Stokes equations' data is the closest approximation to the corresponding natural phenomenon.
For the case in hand, data sets obtained from direct numerical simulations are the best possible approximation to real-world data of reversals, and capture their dynamics fully without any low-dimensional approximation.
Now, the question is: can these irregular time series be a chaotic solution of some simpler LDM?
If that is indeed the case, then one can use the technique of delay embedding (originally proposed by \citet{packard80}, and formalized later by \citet{takens81} and others \cite{aeyels81,mane81,sauer91}) to compute certain invariants, e.g., the dimension, of the underlying system.
If the computed dimension is finite, and preferably low, one can conclude that it is more appropriate to model the time series as a chaotic solution of an LDM\@.
In other cases, a very high-dimensional, or even a simpler stochastic model would be more appropriate.

The ideas and techniques described above fall under the broader domain of nonlinear time series analysis and they have been used over the last three decades to find low-dimensional behavior in many natural phenomena, including turbulence~\cite{kantz04}.
However, they have not always been successful, and some of the previous studies, especially those suggesting the existence of climatic attractors, have been met with skepticism~\cite{ruelle90}.
Today, the general consensus~\cite{grassberger91,kantz04} seems to be that nonlinear time series techniques may not be very reliable when short and nonstationary data sets are used (see Ref.~\cite{mannattil16} for a practical example).
This is one reason why we are using long noise-free data sets obtained from direct numerical simulations in our study.
We further remark that such an analysis only reveals whether or not an LDM is admissible for a certain physical process.
Other matters, e.g., how and why the process displays low-dimensional behavior, extracting an LDM from a time series, etc., are outside the purview of nonlinear time series analysis and requires deeper study.

Using this nonlinear analysis, we find evidence of low-dimensional behavior in zero-Pr flow reversals, opening the possibility of modeling them using an LDM\@.
However, we fail to find similar evidence of low-dimensional behavior in infinite-Pr reversals.
Hence, low dimensionality is not self-evident, and it cannot be deduced solely from numerical simulations or experiments unless one performs relevant nonlinear analysis of the resulting data.
Thus, our results question the generally prevalent assumption that flow reversals are low dimensional.
Our findings also highlight the difficulty in formulating a generic LDM for flow reversals in fluid convection.
In short, one should attempt to construct different models for flow reversals occurring in different fluid systems.
These are the main results of this paper.


\section{Numerical simulations}
\label{sec:numerics}

Many researchers have explored the idealized setup of convection, viz., Rayleigh--B\'enard convection (RBC) where the fluid is confined between two parallel horizontal plates and is heated from below~\cite{chandra61,jkb87, Verma:NJP2017}.
The governing equations for RBC under the Oberbeck--Boussinesq approximation are,
\begin{subequations}
  \begin{align}
    \frac{\partial \mathbf{u}}{\partial t} + (\mathbf{u} \cdot \nabla) \mathbf{u} &= -\frac{\nabla \sigma}{\rho_0} + \alpha g \theta \mathbf{\hat{z}} + \nu \nabla^2 \mathbf{u}, \label{eq:rbc_1}\\
    \frac{\partial \theta}{\partial t} + (\mathbf{u} \cdot \nabla) \theta &= \frac{\delta T}{l}u_z + \kappa \nabla^2 \theta, \label{eq:rbc_2}\\
    \nabla \cdot \mathbf{u} &= 0.\label{eq:rbc_3}
  \end{align}
\end{subequations}
Here, $\mathbf{u}$, $\sigma$, $\theta$, $\alpha$, ${g}$, $\nu$, and $\kappa$ are respectively velocity field, pressure field, temperature fluctuation about the conduction profile, thermal expansion coefficient, acceleration due to gravity, kinematic viscosity, and thermal diffusivity.
Also, the Prandtl number $\mathrm{Pr} \equiv \nu/\kappa$ and the Rayleigh number $\mathrm{Ra} \equiv g\alpha l^3(\delta T)/\nu\kappa$, where $\delta T$ is the temperature difference between the fixed bottom and top plates kept at a distance $l$ apart.
As the Rayleigh number (which is a measure of the strength of the destabilizing buoyancy force relative to the stabilizing viscous force in the fluid) crosses a critical value, depending on specific boundary conditions, heat transfer via convection ensues.

Researchers~\cite{niemela01,sreeni02,brown05,xi06,sugiyama10,chandra11,chandra13} have studied flow reversals in RBC for moderate Pr, especially for air ($\mathrm{Pr}\approx 0.7$) and water ($\mathrm{Pr}\approx 7$).
\citet{sugiyama10} have experimentally investigated flow reversals in a quasi-two-dimensional cell for $\mathrm{Ra}$ ranging from $10^7$ to $10^{10}$, and have compared the results with numerical simulations in a two-dimensional rectangular geometry.
The time intervals between successive reversals in such systems vary from $10$ to $10^4$ eddy turnover times depending on the $\mathrm{Ra}$ used.
Hence, the number of reversals in such systems are limited when the $\mathrm{Ra}$ is large.
Even fewer reversals are recorded in three-dimensional computer simulations~\cite{mishra11}.
Furthermore, as \citet{sugiyama10} showed, for certain values of Pr and Ra, reversals cannot even be observed, despite long observation and/or simulation times.
Since one requires a large number of flow reversals for reliable nonlinear analysis, in this paper, we focus on $\mathrm{Pr}=0$ and $\infty$ for which the number of reversals at appropriate $\mathrm{Ra}$ is large.
RBC systems with zero and infinite Prandtl numbers are of practical importance as well, as they represent convections in liquid metals and the Earth's mantle respectively.

For both infinite and zero Prandtl numbers, we use the pseudospectral code \textsc{Tarang}~\cite{Verma:Pramana2013} to simulate Eqs.~\eqref{eq:rbc_1}--\eqref{eq:rbc_3}.
These equations are time stepped using a fourth-order Runge--Kutta scheme, and the fields are dealiased using the $2/3$ rule.
For very large Prandtl numbers, one can use $\kappa \sqrt{\mathrm{Ra}}/l$, $l$, and $\delta T$ as the velocity, the length, and the temperature scales respectively, in order to write Eqs.~\eqref{eq:rbc_1}--\eqref{eq:rbc_3} in the following nondimensionalized form:
\begin{subequations}
  \begin{align}
    \frac{1}{\mathrm{Pr}} \left[\frac{\partial \mathbf{u}}{\partial t} + (\mathbf{u} \cdot \nabla)\mathbf{u} \right] &= - \nabla \sigma + \theta \hat{z} + \frac{1}{\sqrt{\mathrm{Ra}} } \nabla^2 \mathbf{u}, \label{eq:u_non_inf} \\
    \frac{\partial \theta}{\partial t} + (\mathbf{u} \cdot \nabla) \theta &= u_z + \frac{1}{\sqrt{\mathrm{Ra}} } \nabla^2 \theta, \label{eq:th_non_inf} \\
    \nabla \cdot \mathbf{u} &= 0.
    \label{eq:cont_inf}
  \end{align}
\end{subequations}
As $\mathrm{Pr} \rightarrow \infty$, the left-hand side of Eq.~(\ref{eq:u_non_inf}) vanishes and consequently, $\theta$ and $\mathbf{u}$ are linearly related.
Thus, we only need to solve Eq.~(\ref{eq:th_non_inf})~\cite{Pandey:PRE2014}.
Since the flow profile of infinite-Pr RBC is quasi-two-dimensional~\cite{schmalzl04,poel13,Pandey:Pramana2016}, we can simulate it quite accurately in a two-dimensional geometry.
The two dimensionalization has an added advantage that we can run our simulations for much longer times, and thus obtain a large number of reversals required for proper nonlinear analysis.
Pandey \emph{et al.}~\cite{Pandey:PRE2014,Pandey:Pramana2016} also showed that the flow behavior for $\mathrm{Pr}=100$ and beyond is quite similar to that of infinite-Prandtl number RBC\@.
Thus, infinite-Pr RBC is also a good representative of RBC with large Pr.

We simulate infinite-Pr RBC in a two-dimensional closed box of unit aspect ratio with $256^2$ grids for $\mathrm{Ra} = 10^8$~\cite{verma15} and run our simulation for $3.5 \times 10^5$ time units.
(In this unit $l^2/(\kappa \sqrt{\mathrm{Ra}}) = 1$.)
This yields about $750$ flow reversals.
These reversals are observed at $(x, z) = (0.0625, 0.500)$ in the aforementioned box.
For the velocity field we employ the free-slip boundary conditions on both vertical and horizontal walls, and for the temperature field we employ conducting and insulating boundary conditions at the horizontal and the vertical walls respectively.
This reversal dynamics has been analyzed in detail and it has been shown that the dominant single-roll structure vanishes briefly during a reversal in preference to several secondary structures~\cite{verma15}.
After a reversal, the dominant single-roll structure reappears but with an opposite sense of rotation.
\begin{figure*}
  \centering
  \includegraphics[]{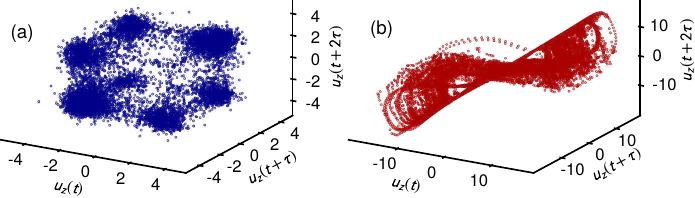}
  \caption{Three-dimensional phase portraits of the phase space reconstructed from the time series corresponding to the vertical real space velocities $u_z(t)$ for (a)~infinite Pr and (b)~zero Pr.}\label{fig:ps}
\end{figure*}

For Prandtl numbers close to zero, we can use $l$, $\nu/l$, and $\mathrm{Pr} (\delta T)$ as the length, the velocity, and the temperature scales respectively to arrive at:
\begin{subequations}
  \begin{align}
    \frac{\partial \mathbf{u}}{\partial t} + (\mathbf{u} \cdot \nabla) \mathbf{u} &= - \nabla \sigma + \mathrm{Ra} \theta \hat{z} + \nabla^2 \mathbf{u}, \label{eq:u_zero} \\
    \mathrm{Pr}\left[\frac{\partial \theta}{\partial t} + (\mathbf{u} \cdot \nabla)\theta\right] &= u_z + \nabla^2 \theta, \label{eq:th_zero} \\
    \nabla \cdot \mathbf{u} &= 0.
  \end{align}
\end{subequations}
At $\mathrm{Pr} = 0$, Eq.~(\ref{eq:th_zero}) reduces to a linear equation. Thus, the only nontrivial equation to simulate is Eq.~(\ref{eq:u_zero}), thus reducing the computational complexity.  Additionally, it has been observed~\citep{pal09} that the flow becomes turbulent at relatively small Rayleigh numbers due to inherent instability of the system.  Hence, we obtain turbulent signal with reversal near the convection onset, which can be simulated with a smaller grid resolution.  The reduced computational complexity and the fewer number of grids again help us obtain a larger number of reversals in a given simulation time.
Thus, we simulate zero-Pr RBC near the convection onset with $\mathrm{Ra} = 690$ in a three-dimensional box of dimension $2\sqrt{2}:2\sqrt{2}:1$ and $64^3$ grids.
We employ perfectly conducting and free-slip boundary conditions at the top and bottom plates, and periodic boundary conditions along the horizontal directions.
In the total simulation time of $3 \times 10^3$ momentum diffusion times ($l^2/\nu$), we observe about $400$ reversals at $(x, y, z) = (\sqrt{2}, \sqrt{2}, 0.09375)$ in the bulk of the fluid.
Just like the infinite-Pr reversals these reversals also appear random, but, as we show in the next section, they are fundamentally different in nature.


\section{Nonlinear time series analysis}

For both infinite and zero $\mathrm{Pr}$, we investigate the time series of the velocity fields at many points in the fluid as well as that of several Fourier modes.
For the sake of concreteness and in order to avoid cluttering the paper with qualitatively similar results, we shall focus only on four time series---two each for infinite and zero Pr, viz., vertical velocity component at a point in the bulk of the flow and its highest energy containing Fourier mode.
The mode happens to be $(1,0,1)$ and $(1, 1)$ depending on whether the simulation is three-dimensional or two-dimensional.
We select these modes because we anticipate such modes to be dynamically the most important ones in determining the flow.
In our simulations of the infinite-Pr reversals, we use a free-slip boundary condition.
Thus, the Fourier modes only possess real components with the temporal evolution of the most energetic mode [i.e., $(1,1)$] being very similar to that of the vertical velocity of the fluid near the side walls~\cite{verma15}.
However, for zero-Pr reversals, the Fourier modes possess both complex and real components.
It has been  shown that during a flow reversal, the complex phase of the first Fourier mode  jumps abruptly by a significant amount, and quite often by about $180{^\circ}$~\cite{mishra11}.
Thus, even though the real or the imaginary parts of the Fourier mode are not completely correlated with the real-space velocity field, they exhibit statistically similar reversal properties.
It should also be remarked that the $(1,0,1)$ mode is one of the most energetic modes as well.
Thus, for both infinite and zero Pr we only use the real part of the Fourier modes for the analysis.
We sample the time series for infinite and zero Pr at intervals of 2 and 0.01 time units and pick segments of lengths $N \approx 160,000$ and $N \approx 260,000$ points respectively.


\subsection{Delay vector}
\label{sub:delay}

Let any one of the aforementioned time series be represented by $\{x_i\}$.
One can construct $d$-dimensional delay vectors $\{\mathbf{y}_i^{(d)}\}$ from time delayed values of $x_i$
\begin{equation}
  \mathbf{y}_i^{(d)} \equiv (x_i, x_{i + \tau}, x_{i + 2\tau},
    \ldots, x_{i + (d - 1)\tau})
\end{equation}
where $d$ is the embedding dimension and $\tau$ is the time delay.
With a time series of length $N$, this would result in $M = N - (d-1)\tau$ delay vectors.
Under generic conditions, the delay embedding theorems guarantee that certain invariants (e.g., Lyapunov exponents, Kolmogorov--Sinai entropy, correlation dimension, etc.)~of the underlying attractor are preserved in the reconstructed phase space containing $\{\mathbf{y}^{(d)}_i\}$ for dimensions $d > 2D_B$.
Here $D_B$ is the box counting dimension of the original attractor.
We emphasize that this condition is only a sufficient condition and there may exist a minimum embedding dimension $d_\text{min} < 2D_B$, at which the attractor fully unfolds.
\begin{figure*}
  \centering
  \includegraphics[]{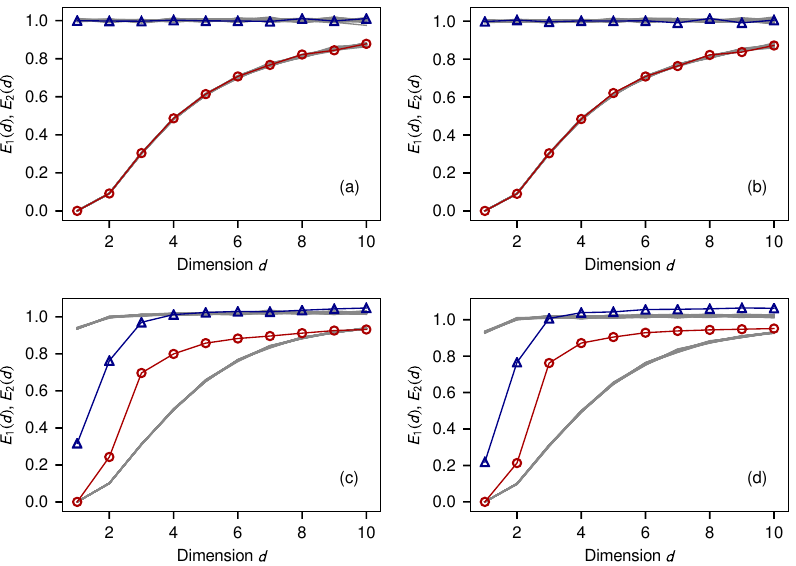}
  \caption{$E_1(d)$ (red lines with circles) and $E_2(d)$ (blue lines with triangles) curves for the vertical component of the velocities for (a)~infinite Pr and (c)~zero Pr, and their largest energy containing Fourier modes [(b) and (d), respectively].  See Eqs.~(\ref{eq:E1}) and (\ref{eq:E2}) for definitions of $E_1(d)$ and $E_2(d)$.   For the infinite-Pr data, the corresponding curves for the 39 surrogates (solid gray) coincide with the main curves, whereas for zero-Pr data the curves for the surrogates are significantly different.  In each of the panels, the lower and the upper gray curves correspond respectively to $E_1(d)$ and $E_2(d)$ for the relevant surrogate series.}\label{fig:afn}
\end{figure*}

Theoretically, for embedding a chaotic attractor, any nonzero $\tau$ can be used.
Nevertheless, for data with a finite number of points, one has to choose an appropriate $\tau$ so that the components of the delay vectors stay independent and the attractor unfolds properly.
A very small value of $\tau$ would result in highly correlated components and an attractor lying along the phase space diagonal, which can lead to spurious results.
At large values of $\tau$, the components lose all dynamical correlations and the reconstruction fails to represent the underlying dynamics.
This is often called ``overfolding'' and the resultant attractor is a structureless collection of points spread all over the phase space~\cite{galka00}.

Two popular choices for $\tau$ are the locations of the first minimum of the mutual information between $\{x_i\}$ and $\{x_{i+\tau}\}$, and of the first zero of the autocorrelation function of $\{x_i\}$~\cite{kantz04}.
For the four time series at hand, neither of these choices work.
For the infinite-Pr time series [Fig.~\ref{fig:series}(a)], there are no well-defined minima in the mutual information curves and the autocorrelation functions do not vanish for any reasonable $\tau$.
Therefore, we have employed another frequently used choice: the autocorrelation time ($\approx 200$).
The three-dimensional phase portrait of the phase space reconstructed from the infinite-Pr time series using this time delay is showcased in Fig.~\ref{fig:ps}(a).
Unlike the infinite-Pr reversals, the zero-Pr reversals occur as quick short jumps [Fig.~\ref{fig:series}(b)] resulting in time series with most points concentrated near their means.
In order to resolve these short but numerous reversals, we have sampled the time series with a high temporal resolution.
This inadvertently makes the autocorrelation time high ($\approx 400$).
But we lose information about the short reversals if we use such a large $\tau$.
We have found out that a more appropriate choice for $\tau$ here is the statistical mode of the waiting times (i.e., the time intervals between the reversals), which is approximately 20.
This is somewhat analogous to choosing the quarter of the average time period as the time delay for series with a periodic component~\cite{kantz04}.
We also remark that methods that rely on the inherent geometry of the reconstructed attractor, such as the one by~\citet{rosen94}, give similar results.
Using this time delay, we obtain a properly unfolded attractor featuring two prominent lobes as shown in Fig.~\ref{fig:ps}(b).
On the other hand, using larger time delays (such as the autocorrelation time) results in an overfolded attractor.


\subsection{Averaged false neighbors method}

\citet{kennel92} has described a method in which $d_\text{min}$ is picked to be the dimension at which the fraction of false nearest neighbors in the reconstructed attractor goes below a certain threshold.
By false nearest neighbors, we mean points that are close together solely because of trajectory crossings that occur when we embed the attractor in a lower-dimensional phase space.
The distance between such points grow large when we increase the embedding dimension, indicating that the chosen embedding dimension is not good enough to reconstruct the attractor.
Unfortunately, this method involves certain subjective parameters and the results often depend on them.
The averaged false neighbors (AFN) method~\cite{cao97} was created to overcome this issue.
The average magnification $E(d)$ of near-neighbor distances when one goes from dimension $d$ to $d + 1$ is given by
\begin{equation}
  \begin{aligned}
    E(d) = \frac{1}{N - \tau d} \sum^{N - \tau d}_{i = 1} \frac{\|\mathbf{y}_i^{(d + 1)} - \mathbf{y}_{n(i,d)}^{(d + 1)}\|} {\|\mathbf{y}_i^{(d)} - \mathbf{y}_{n(i,d)}^{(d)}\|}.
  \end{aligned}
\end{equation}
Here $\|\cdot\|$ is the Euclidean norm and $\mathbf{y}_{n(i,d)}$ is the nearest neighbor of the $i$th time-delay vector $\mathbf{y}_i$ in the reconstructed $d$-dimensional phase space.
Note that we are computing the magnification $E(d)$ by averaging the magnification of near-neighbor distances over all points.
This the reason why this procedure is called the averaged false neighbors method.
We also enforce a minimum temporal separation (equal to the autocorrelation time of $\{x_i\}$) between the near neighbors so that the method does not report spurious low dimensions due to serial correlations~\cite{theiler86,fredkin95}.
It is seen that for deterministic time series, the ratio
\begin{equation}
  E_1(d) = \frac{E(d + 1)}{E(d)}
  \label{eq:E1}
\end{equation}
saturates after $d_\text{min}$, when all the false nearest neighbors have been eliminated.
On the other hand, for stochastic time series, $E_1(d)$ never saturates with $d$.
But owing to the finiteness of data, $E_1(d)$ stops changing after a certain $d$.
Therefore, we compute
\begin{equation}
  \begin{aligned}
    E^*(d) = \frac{1}{N - \tau d} \sum^{N - \tau d}_{i = 1} |x_{i + \tau d} - x_{n(i,d) + \tau d}|,
  \end{aligned}
\end{equation}
which represents the average of the absolute difference between the components that get added to the near neighbors while moving up by a dimension.
For the case of pure noise, components that get added would be picked randomly from the underlying probability distribution.
Hence, we expect
\begin{equation}
  E_2(d) \equiv \frac{E^*(d + 1)}{E^*(d)} \approx 1.
  \label{eq:E2}
\end{equation}
But for deterministic data we would expect at least one $d$ for which $E_2(d) \ne 1$.
\begin{table*}
  \caption{Results of surrogate analysis using the variation coefficient $V_{E_2}$ of $E_2(d)$ as the test statistic for various time series. $V_{E_2}$ of the original time series, along with the minimal and maximal values of $V_{E_2}$ for the corresponding surrogate data sets is given in each case.  The null hypothesis is rejected if $V_{E_2}$ of the original data set falls outside this range.  If $V_{E_2}$ falls inside this range, the percentile indices localizing it in the $V_{E_2}$ distribution of the surrogate series is also reported.}
  {\label{tab:surr}}
  \begin{ruledtabular}
  \begin{tabular}{lcccl}
    Time series & $V_{E_2}$ & Surrogates $V_{E_2}$ $\left[\text{min}, \text{max}\right]$ & Percentile indices & Null hypothesis\\
    \hline
    zero-Pr probe data & 0.236 & $\left[0.0232, 0.0262\right]$ & \ldots & Rejected\\
    zero-Pr mode data & 0.271 & $\left[0.0243, 0.0283\right]$ & \ldots & Rejected\\
    infinite-Pr probe data & 0.00511 & $\left[0.00295, 0.00946\right]$ & 57--58  & Not rejected \\
    infinite-Pr mode data & 0.00647 & $\left[0.00304, 0.00794\right]$ & 92--93 & Not rejected\\
  \end{tabular}
  \end{ruledtabular}
\end{table*}

Figures~\ref{fig:afn}(a) and~\ref{fig:afn}(b) show the $E_1(d)$ and $E_2(d)$ curves for the infinite-Pr time series.
The $E_1(d)$ curves for both the infinite-Pr time series keep rising steadily with the embedding dimension without any saturation, resembling a stochastic system.
We can also see that the $E_2(d)$ values are very close to $1$ for all $d$.
Note however that we obtained the infinite-Pr time series using purely deterministic direct numerical simulations.
Thus, although the time series exhibits characteristics \emph{imitating} a stochastic system, the underlying dynamics is a very high-dimensional deterministic one to be precise.
In comparison, the results [Figs.~\ref{fig:afn}(c) and~\ref{fig:afn}(d)] for the zero-Pr time series are distinctly different: $E_1(d)$ saturates for $d \gtrapprox 5$ and $E_2(d) \neq 1$ at many $d$, telling us that the underlying dynamics is low dimensional.
Note that $d_\text{min} = 5$ is also consistent with the criterion that dimension estimates larger than $2\log_{10}{M}$ ($\approx 10$ for the zero-Pr time series at $d=10$) cannot be justified while working with $M$ delay vectors~\cite{eckmann92}.

In principle, we can use any other test of determinism to distinguish these series.
We have chosen the AFN method primarily because of its simplicity and also because it has arguably less subtleties and pitfalls when compared to other methods, e.g., computing the correlation dimension~\cite{mannattil16}.
Furthermore, the quantity $E_2(d)$ has been shown to be a powerful indicator of determinism and can accurately distinguish correlated noisy time series from deterministic ones---something that is often difficult to do with other methods~\cite{ramadani07}.


\subsection{Surrogate analysis}

Tests for determinism often give spuriously low dimension estimates when performed on short noisy time series.
A notorious case is that of correlated noise, which may appear to be low dimensional because of temporal correlations.
For this reason, one usually performs surrogate analysis to verify whether the observed low dimensionality is due the presence of nonlinearity in the time series.
In surrogate analysis~\cite{theiler92} one creates linearly correlated random data sets that mimic certain characteristics (e.g., power spectrum, distribution, etc.)~of the original time series.
A null hypothesis that the surrogates are sufficient to explain the results of the nonlinear analysis is assumed.
The determinism tests are then carried out on the surrogates and the null hypothesis is rejected only if there is a significant difference in the results.
A very general, yet testable null hypothesis is that the time series is correlated Gaussian noise.
Any non-Gaussianity in the time series is assumed to be the result of a static monotonic transformation of an originally Gaussian series~\cite{schreiber96}.
For each time series we use in our analysis, we generate 39 surrogate data sets satisfying this null hypothesis.
We use the variation coefficient $V_{E_2}$ of $E_2(d)$ (i.e., the ratio of the standard deviation of $E_2(d)$ to its mean) as the test statistic~\cite{ramadani07}.
We then compare the $V_{E_2}$ values of the original time series with the values obtained from the corresponding surrogate data sets using a two-sided rank based test, which has a significance level of $0.05$ when 39 surrogates series are used~\cite{kantz04}.
\begin{figure*}
  \centering
  \includegraphics[]{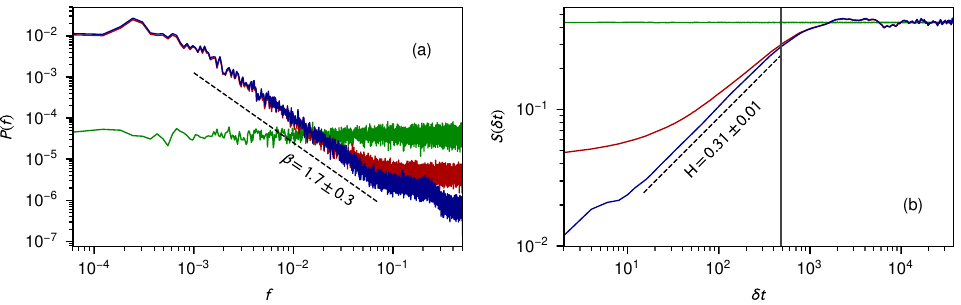}
  \caption{(a)~Power spectra $P(f)$ of Fourier mode $(1, 1)$ for the infinite-Pr reversals (blue), synthetic data set A (green), and synthetic data set B (red).  See the main text for details regarding the synthetic data sets.  (b)~Second-order structure functions $S(\delta t)$ for all the three data sets.  The color code is same as in~(a).  The gray vertical line indicates the average waiting time of the reversals.}\label{fig:hurst}
\end{figure*}

Table~\ref{tab:surr} summarizes the results of the surrogate analysis, where $V_{E_2}$ values for the original time series and its ranges for the surrogate data sets are given.  If $V_{E_2}$ of the original data set falls inside this range, we cannot reject the null hypothesis.
It is clear from Table~\ref{tab:surr} that the null hypothesis of linearly correlated noise can be rejected at a significance level of $0.05$ for both the zero-Pr time series, for which the $V_{E_2}$ values are much higher than the values obtained for the surrogates.
This can also be qualitatively seen from the gray $E_1(d)$ and $E_2(d)$ curves of the surrogates presented in Figs.~\ref{fig:afn}(c) and~\ref{fig:afn}(d), which deviate significantly from the corresponding curves for the original time series.
Thus, this confirms that the low dimensionality we have observed in zero-Pr reversals is due to the presence of nonlinearity.
On the other hand, we cannot reject the null hypothesis for either of the infinite-Pr time series.
Each of the surrogate time series replicates the results of the AFN algorithm rather well.
Thus, as far as the AFN method is concerned, this data is no different from correlated noise.
Since power spectra and structure functions are better suited to study noise-like data, we explore them in the next section.


\section{Spectral analysis of infinite-{Pr} reversals}

Power spectrum $P(f)$ of the Fourier mode $(1, 1)$ for the infinite-Pr reversals shows a $1/f^{\beta}$-like dependence at intermediate frequencies.
One can estimate the spectral exponent $\beta$ by computing the mean of the local slopes $\mathrm{d}\,(\log{P(f)})/\mathrm{d}\,(\log{f})$ and can take their standard deviation to be the error in $\beta$.
For the infinite-Pr reversals this gives us $\beta = 1.7 \pm 0.3$.
A similar power law dependence is also seen in the autocorrelation function of the time series, and such a series is said to possess ``long-range correlations''.
One can also quantify long-range correlations in a time series by using a second-order structure function $S(\delta t)$, which scales as
\begin{equation}
  S(\delta t) = \langle |x_{i + \delta t} - x_i|^2 \rangle \propto \delta t^{2H}.
\end{equation}
The scaling parameter $H$ quantifies the correlations between the past and future increments of the time series, with the $H < 0.5$ indicating negative correlations and $H > 0.5$ indicating positive correlations~\cite{feder88}.
$S(\delta t)$ for the infinite-Pr time series [Fig.~\ref{fig:hurst}(b)] has a clear linear scaling range for intermediate time scales with an $H = 0.31 \pm 0.01$.
Hence, the infinite-Pr time series possesses long-range anti-correlations.
Thus, on an average, increasing trends in the time series are followed by decreasing trends (and vice-versa), which is exactly what one would intuitively expect in a time series exhibiting reversals.
Over timescales larger than the average waiting time of the reversals, the data becomes stationary and the structure function becomes flat.

In order to verify that these correlations are due to the reversals (jumps) themselves, we make use of two synthetic data sets obtained using $x_i' = x_j$ (Set A) and $x_i' = \mathrm{sgn}(x_i)|x_j|$ (Set B), where $j$ is a random integer drawn from $[1, N]$ without replacement.
Though both data sets have been randomized and their local correlations fully destroyed, Set B changes its sign in synchrony with the original series thereby exhibiting ``synthetic reversals''.
Power spectrum and structure functions for these synthetic data sets are also presented in Fig.~\ref{fig:hurst}.
As expected, the power spectrum for Set A shows equal power at all frequencies and the corresponding structure function is flat.
On the other hand, the power spectrum for Set B has a $1/f^{\beta}$-like dependence with $\beta = 1.5 \pm 0.3$ and the structure function has a nontrivial scaling.
Although we do not recover the scaling relations seen in the original time series exactly, it is evident that they are predominantly brought forth by the sequence of jumps corresponding to the reversals.

$1/f^\beta$-like spectra is ubiquitously seen in several natural systems~\cite{west89}.
In particular, other field reversals, e.g., geomagnetic field reversals, are also known to display similar antipersistence, $1/f^{\beta}$-like behavior, and long-range correlations~\cite{pelletier99,dmitruk14}.
Although low-dimensional chaotic systems often have broadband spectra~\cite{schuster06}, they can only generate $1/f^\beta$-like spectrum over a limited range of frequencies~\cite{theiler91}.
This could be the reason why we failed to find low-dimensional behavior in the infinite-Pr data, which exhibits $1/f^{\beta}$-like behavior over two decades (which as we saw above is mainly due to the reversals).


\section{Discussions and conclusion}

Field reversals are some of the most important problems of geophysics and astrophysics, and several simple LDMs have been proposed to describe them.
Using techniques from nonlinear time series analysis, we ask whether such elementary descriptions are always appropriate for flow reversals in Rayleigh--B\'{e}nard convection.
Though we have found evidence of low-dimensional behavior in reversals occurring in zero-Pr convection, we have failed to find similar signatures in infinite-Pr reversals, despite it being generated by deterministic direct numerical simulations.
It follows that a rudimentary low-dimensional description is suitable only for zero-Pr reversals, and not for infinite-Pr reversals.
However, instead of using a high-dimensional deterministic model to capture the dynamics of infinite-Pr reversals, it could perhaps be more convenient to model them using a simpler stochastic model.

Let us now try to understand these contrasting results.
Apart from Pr, the infinite-Pr and zero-Pr systems differed in other parameters as well.
Thus, one cannot simply assume that a change in Pr is what caused these systems to behave differently.
\citet{paul11} showed that in RBC, the transition Rayleigh number for getting turbulence increases with increasing Prandtl number.
For example, for $\mathrm{Pr}=6.8$, turbulence first appears near $\mathrm{Ra} \approx 3.3\times10^4$, which is much larger than the critical Ra for zero or small Prandtl numbers where transitions to turbulence occur near the onset of convection itself.
This in turn leads to the excitation of a large number of Fourier modes at large Pr.
Though reversals at zero and infinite Pr are both governed by the same basic laws, at zero Pr the advective term $(\mathbf {u} \cdot \nabla) \mathbf{u}$ of the momentum equation (Eq.~\ref{eq:u_zero}) dominates all the other terms.
For infinite-Pr reversals, however, the dominating term is the $(\mathbf {u} \cdot \nabla) \theta$ term from the temperature equation (Eq.~\ref{eq:th_non_inf}).
Note that it is the $(\mathbf {u} \cdot \nabla) \theta$ term that is responsible for generating higher Fourier modes.
Consequently, in infinite-Pr reversals, a large number of modes are excited, making it resemble a stochastic system.

Likewise, the number of active modes are limited in the case of zero-Pr reversals, consistent with the low dimensionality we have observed.
We have also seen that the zero-Pr time series can be properly embedded in a $5$-dimensional phase space, whose dimension is less than the $13$ modes used in an LDM proposed for these reversals by~\citet{pal09}.
Though this model is quite successful in capturing several features of zero-Pr convection (e.g., bifurcations), given the above results, one can perhaps use even fewer Fourier modes and devise a simpler model, e.g., with $5$ or $6$ modes.
One can perhaps also ask if such a low-dimensional model for zero-Pr reversals can be modified by adding a noise term to capture the dynamics of infinite-Pr reversals.
The analysis techniques used in the present study will only enable us to compare the invariants of the attractors reconstructed from the actual series and the modified model.
Since attractors of different physical systems can often have similar invariants, we unfortunately would not be able to derive a stronger conclusion from such a comparison.

The results presented in this paper are entirely based on data obtained from direct numerical simulations of the Navier--Stokes equations adapted for the convection problem.
Choosing to work with zero- and infinite-Pr reversals, which closely resemble convections in the Earth's mantle and liquid metals, has enabled us to generate long enough time series, while also presenting us with physically relevant data.
Using numerical data has also made it possible for us to accurately identify the contrasting behaviors of these reversals without worrying about additional complications such as observational noise.
However, unlike the data we have used in our study, most real-world time series of field reversals tend to be short.
For instance, only less than 50 reversals were observed by~\citet{berhanu07} in an experimental dynamo containing liquid sodium.
And in the case of geomagnetic reversals, we only know about the exact sequence of reversals, making conventional nonlinear analysis impossible, despite having records close to 100 million years~\cite{lowrie07}.
Similarly, consider the regular 11-year sunspot cycle, which is a direct consequence of solar magnetic field reversals.
Though there have been previous attempts to detect low-dimensional behavior (e.g., in the form of periodic orbits) in sunspot activity, an increasingly prevalent view is that there is no strong evidence for it, given the shortness of available data~\cite{petrovay10}.

In summary, we have shown that the dynamics of flow reversals in RBC can be quite complex.
For some parameters, the dynamics is low dimensional, but for some others it is not.
This shows that low dimensionality is not obvious, and writing down a generic model for convective flow reversals is probably not very feasible.
It would be interesting to carry out similar studies on reversals occurring at moderate Prandtl numbers such as those corresponding to air and water, some of which have already been studied using experiments and numerical simulations.
However, as we discussed in Section~\ref{sec:numerics}, the durations of these simulations, especially in three dimensions, are prohibitively long to generate time series required for effectively employing methods of nonlinear time series analysis.
Here, surprisingly, experimental data may come to the rescue, e.g., close to 2300 reversals have been observed in certain convection experiments at these Prandtl numbers~\cite{sreeni02}.
Flow reversals have continued to received attention over the years, and newer LDMs~\cite{podvin15} are reproducing their features even more accurately.
However, our conclusions about the difficulty in formulating a single LDM for flow reversals across the entire possible ranges of the parameters are likely to stay.


\subsection*{Acknowledgments}

The authors thank Anindya Chatterjee, Dhrubaditya Mitra, Ishan Sharma, and Pankaj Wahi for fruitful discussions.
M.K.V.~thanks IFCPAR/CEFIPRA for financial support.
S.C.~acknowledges financial support through the INSPIRE faculty fellowship (DST/INSPIRE/04/2013/000365) conferred by the Indian National Science Academy (INSA) and the Department of Science and Technology (DST), India.


%

\end{document}